\documentstyle[multicol,aps,prl,psfig]{revtex}

\begin{document}
\twocolumn[\hsize\textwidth\columnwidth\hsize\csname 
 @twocolumnfalse\endcsname

\title{Self-structuring of Granular Media
under Internal Avalanching}
\author{S. Krishnamurthy $^{a}$, V. Loreto $^{a}$,
 H.J. Herrmann $^{a,b}$, S.S. Manna $^{c}$ and S. Roux $^{d}$}
\pagestyle{myheadings}
\address{
$a)$ P.M.M.H. Ecole Sup\'erieure de Physique et Chimie Industrielles, \\
10, rue Vauquelin, 75231 Paris CEDEX 05 France \\
$b)$ ICA1, Univ. Stuttgart, Germany \\
$c)$  Satyendra Nath Bose National Centre for Basic Sciences, \\
Block-JD, Sector III, Salt Lake, Calcutta 700091, India. \\
$d)$ Laboratoire Surface du Verre et Interfaces, Unit\'e Mixte de Recherche
CNRS/Saint-Gobain, \\
 39, Quai Lucien Lefranc, F-93303 Aubervilliers Cedex, France.}

\maketitle
\date{\today}
\begin{abstract}
We study the phenomenon of internal avalanching within the
context of recently proposed ``Tetris'' lattice models for granular
media. We define a recycling dynamics under which the system reaches a 
steady state which is {\em self-structured}, {\em i.e.} 
it shows a complex interplay between textured internal 
structures and critical avalanche behavior. 
Furthermore we develop a general mean-field theory 
for this class of systems and discuss possible scenarios
for the breakdown of universality. 

\end{abstract}
\smallskip
\vskip2pc]

There has been a lot of interest in understanding the internal
structure and geometry of granular packings \cite{grain}. The rich
phenomenology observed in experiments studying compaction,
segregation and force distributions, amongst other things, has
prompted a number of numerical and analytical 
studies. In another context the interest in granular
media has also been triggered by the search for self-organised
criticality (SOC)\cite{soc}. As a result surface avalanches in granular 
piles  have been extensively studied both experimentally \cite{exp} 
and in computer models \cite{soc,sandpiles} in order to identify clear
regimes of  SOC-like behavior \cite{oslo}. In this Letter, we focus
our interest on the interplay between the
internal structure of packings and power-law avalanche distributions. 
We define a steady state dynamics
under which the medium reaches a {\em self-structured critical} 
state. We also focus on the internal structure of this state and find, 
very interestingly, that it can be highly inhomogeneous with strong
segregation  and ordering effects. 

The model we have investigated is
a recently proposed simple lattice model for
describing slow dynamical processes in granular media
\cite{tetris}. The basic ingredients of this model are the
geometric constraints involved in packing particles of different shapes.
This model is seen to reproduce experimentally observed phenomena such as 
slow relaxation in compaction \cite{tetris}, segregation\cite{segtet}, 
as well as aging \cite{nicodemi}. 

Within the context of this model, we study the
phenomenon of internal avalanches occurring under small perturbations.  
But as opposed to previous works\cite{sketal}, 
we focus on the stationary state that a system
reaches under the continued process of removing a particle from the 
bottom layer and adding it back to the top of the system. 
Under this dynamics the system reaches a well defined 
``critical'' steady state in which the avalanche distributions 
decay as power laws. Most interestingly,
we find that in order to achieve this effect, the system restructures
under this dynamics to a very inhomogeneous state with
ordered regions (grains) separated by disordered low-density channels 
(grain boundaries) which
act as preferential pathways for these avalanches. 
We perform our numerical
experiments for particles of several different shapes and find that the
steady state reached is always as described above, with an exponent
for the  power-law distribution which is the same for a large class
of particle shapes.
We furthermore develop a mean-field theory for systems undergoing this 
dynamics and explain, within this context, why we observe a universal
power-law distribution. We elaborate on this point by considering a 
case when an important change in the rules of stability of particles 
changes the steady state reached and hence
the universality class of the phenomenon.

We briefly review the definitions and some basic properties 
of the {\em Tetris} models \cite{tetris} 
used in our simulations. 
Frustration arises in granular packings 
owing to excluded volume effects of 
particles of different shapes.
This geometrical feature is captured in the {\em Tetris} model. 
In the following, we present results for the
simplest version of the model where the particles are either 
rods with two kinds of orientations, more complicated shapes such as
``T''- shaped particles with two kinds of orientations or 
``crosses'' with arms of randomly distributed lengths
in the framework of the so-called {\em Random Tetris Model} (RTM)
\cite{rtm}. 

\begin{figure}[h]
\centerline{
       \psfig{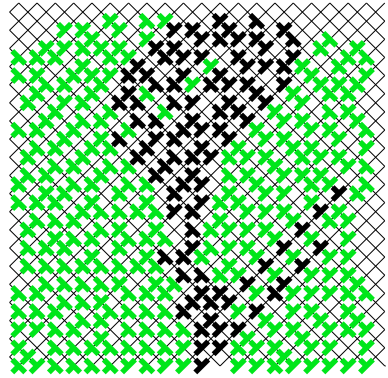}}
        \vspace*{0.1cm}
\caption{A steady state configuration of the {\it Tetris} model 
with ``T-shaped'' 
particles with two different orientations. 
The boundary conditions are periodic 
in the horizontal direction. The black particles are those which 
rearrange in the avalanche caused by removing the lowest particle.}
\label{figtshape}
\end{figure}

The {\em Tetris} model can be defined as a system of 
particles which occupy the sites of a square lattice tilted by $45^0$ 
with periodic boundary conditions in the horizontal direction
(cylindrical geometry) and a rigid wall at the bottom.
Particles cannot overlap and this condition produces
very strong constraints (frustration) on their relative positions. 
This is illustrated for ``T-shaped'' particles in 
Fig.(\ref{figtshape}). 
In general each particle can be schematized
as a cross with arms of different lengths which can be 
chosen in a regular\cite{tetris} or in a random way\cite{rtm}. 
The system is initialized by 
inserting the particles at the top of the system,
one at a time, and letting them move down  under gravity.
The particles perform an oriented random walk on the lattice until 
they reach a stable position defined as a position from which they 
cannot fall any further because of other particles below them. 
The particles retain their orientations as they move, 
{\it i.e.} they are not allowed to rotate.  
We now introduce the following dynamics under which the system evolves. 
A particle is removed from a random position at the base.
This could destabilise its neighbouring particles above
one of which may then fall down  
if the geometry of the packing allows for the motion (i.e., if the 
orientation of the particle fits the local conformation). In this
case, the disturbance propagates upwards
destabilizing particles in the layer above and so on. We update
the system sequentially, moving all the unstable particles down till
the system is once more stable.
The removed particle  is then added back at a
random position at the top of the system. This process is 
continued till the
system reaches a steady state. 

Similar procedures have
been studied before for other models \cite{sny-ball,manna}. While
long-ranged avalanche distributions have been found in 
\cite{sny-ball}, the update rule assumed in \cite{manna} does not lead to
a critical state. We go beyond these previous works
by studying here, in
detail, the interplay between the avalanche distribution and the
density profile of the medium. We explain the means by which the
system reaches a critical state by developing a 
generic mean-field theory for avalanche distributions in dense or
loose packings. We utilise the possibility afforded by this model,
of easily changing particle shapes, to study this behaviour for
a wide variety of particle shapes.
Most interestingly we also find that this ``critical''
steady state is inhomogeneous and strongly ordered, different from
those ordinarily studied in most SOC systems.
These are thus some of the new features reported in the 
present study.

\begin{figure}[h]
\centerline{
       \psfig{figure=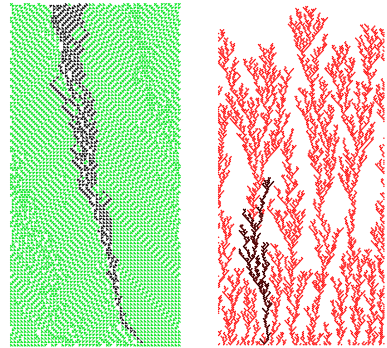,width=5cm,angle=0}}
        \vspace*{0.1cm}
\caption{Typical avalanches in the steady state for a system of 
T-shaped particles {\bf (left)} and {\em sticky} particles 
{\bf (right)}.} 
\label{structure}
\end{figure}

Fig. 2a is a picture of a packing of 
in the steady state. As can be seen
it shows a complex textured structure. 
Namely, beginning from an initial state in which particles of
different shapes are homogeneously mixed,
the packing always ``segregates'' under the dynamics 
so as to form ordered high density grains separated by grain boundaries 
at lower densities. 
All avalanches preferentially propagate inside these grain boundaries
{\it i.e.} no matter where the initial seed, the avalanches
find their way into the boundary region (see Fig.(\ref{structure}-left)). 

\begin{figure}[h]
\centerline{
        \psfig{figure=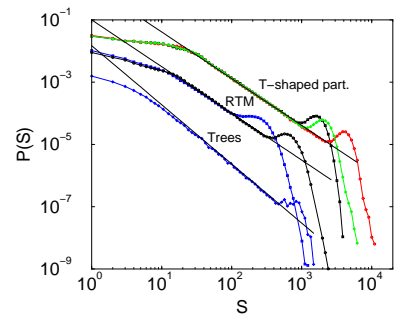,width=5cm,angle=-90}}
        \vspace*{0.1cm}
\caption{$P(s)$ vs $s$
in the steady state for ``T-shaped'' particles and the ``crosses''
({\em RTM}):
$\tau = 1.5 \pm 0.05$ in both cases. The system sizes 
shown are $Lx=100,Ly=500$ and $Lx=200,Ly=650,1000$ respectively for
the ``Tees'' and $Lx=100,Ly=150$ , $Lx=200,Ly=300$ for the
``crosses''. 
The last curve shows the avalanche distribution for
a system of {\em sticky} particles (see text). In this case one gets
$\tau=1.9 \pm 0.05$. The system size shown is $Lx=200,Ly=350$.}
\label{figaval}
\end{figure}

The size of an avalanche is
defined as the total number of particles destabilized by the process of
removing a particle at the bottom.
The size distribution of the avalanches 
decays like a power $P(s) \sim s^{-\tau}$. 
This was studied for the
three different types of particles described above. Time averages were
performed in the steady state over $\sim 10^6$ configurations in order
to obtain good statistics.
Fig.(\ref{figaval}) shows the avalanche distributions obtained 
for two different choices of the particles: the {\em T's}
shown in Fig.(\ref{figtshape}) and particles with random shapes
obtained in the framework of the RTM.
In both cases one observes a scaling behavior for the avalanches
with  $ \tau = 1.5 \pm 0.05$. In the case of the rods, the result is
sensitive to the aspect ratio of the system 
but for systems of about equal width and height
the exponent of the avalanche
distribution is again the same. 

We now turn to a discussion of the avalanche statistics
within the framework of a mean-field theory that we
develop for this class of systems.
It is apparent from our numerical studies reported so far
that under this dynamics,
the system reaches a steady state which is critical. The reason 
could be the
following. Taking out a particle in the last layer creates a void in
the packing. This can either move up (by exchanging place with a
particle), die (if nothing above is destabilized) or free another
neighboring void (and hence multiply) and propagate. The dynamics is
thus essentially like a branching-annihilating process on the
lattice where the probabilities for annihilating $P_0$, branching
$P_2$ and propagation $P_1= 1-P_0-P_2$ depend on the density of the
packing. However there is also a feedback effect. The avalanche
distribution can in its turn affect the density of the system : 
large avalanches that reach the top tend to compactify the system 
and small avalanches make the system looser.  

We can make the above arguments more precise in the following
way. Let $ \rho_h (t)$ be the cumulative 
density of the system up to height $h$ at time $t$. Then
the density of the system at time $t+1$ will be:

\begin{equation}
\rho_h(t+1)- \rho_h(t) = - 1/L^2 + a(t) *h^{\gamma}/L^2
\label{couple1}
\end{equation}
where $L$ is the linear size (height) of the system. The first term
of the RHS represents the effect of removing one particle. 
This is the sole 
contribution of avalanches which die before reaching the
height $h$.
On the other hand, those avalanches which reach 
at least a height $h$ have the additional effect of
increasing the density of the system by an amount equal 
to the number of voids which escape at $h$. 
This is equal to the width of the avalanche at $h$.
If the avalanches are self-affine (as in \cite{sny-ball},
and also in the case studied here), {\it i.e.} an
avalanche of height $h$ has a width of $h^{\gamma}$, then
the density increase is precisely given by the second term. 
The coefficient $a(t)$ is just a random variable which takes
the value $1$ whenever an avalanche reaches at least a height $h$
and a value $0$ otherwise.
In the steady state, we can perform a time average on
Eq.(\ref{couple1}). We expect the LHS to vanish in this case.
To evaluate the RHS, we note that the time
average of $a(t)$ is simply $1- \int_{0}^{h} P(h) dh$.
We measure $P(h)$ numerically and 
observe the existence of a 
scaling region where $P(h) \sim h^{-\beta}$ with $\beta = 1.95 \pm 0.05$. 
We also independently measure $\gamma$ (by measuring
the characteristic size $s^* = h^ {1+\gamma}$ 
of the avalanche size cut-off at height $h$). We find
$\gamma = 0.9 \pm 0.1$.

The above equation makes a prediction for the avalanche exponent.
The steady state condition requiring that the average
density of the system  $\langle \rho \rangle =const$ 
implies that $ \beta -1 = \gamma$.  From the numerically measured
value of $\beta$ mentioned above, we see that we find $\gamma \sim
0.95$ consistent with the numerically measured value of $\gamma$. 
Making a change of variables from
the avalanche height $h$ to the avalanche size $s$
gives us the relation 
$ (1-P_0 (h))h^{\gamma} = h^{\gamma}/h^{(1+\gamma)(1-\tau)} =1$
where $P(s)  \sim s^{-\tau}$ is the avalanche size distribution
in the steady state.
The above scaling relation for $\beta -1 = \gamma$ then
translates to (also obtained in
\cite{sny-ball} using a steady state argument) 
\begin{equation}
\tau = 1+ \gamma/(1+\gamma).
\label{taugam}
\end{equation}
Using again our  numerical estimate for $\gamma$ we find $\tau \sim
1.47 \pm 0.05$ consistent with the data shown in Fig. {\ref{figaval}}.

A more complete and self-consistent description of the 
observed phenomenology can be obtained complementing equation
(\ref{couple1}) with an equation for the avalanche distribution 
$P(s)$ in terms of $\rho$, {\it i.e.} with an equation
\begin{equation}
P(s(t)) = F(\rho(t))
\label{couple2}
\end{equation}
where $F$ indicates a generic function of $\rho(t)$.
The two coupled equations (\ref{couple1}) and (\ref{couple2}) 
should then describe the evolution of the
system to a steady state given by a critical density $\rho_c$ with an
avalanche distribution decaying as a power-law. In
general, it is difficult to write an exact equation
for the avalanche distribution in terms of the
density except for avalanches propagating on the Bethe lattice \cite{szkl}.
In this case, it is possible to show quite simply that the
feedback effect of Eq.(\ref{couple2}) on Eq. (\ref{couple1})
results in the system reaching a critical density $\rho_c$
given simply by the equation $ P_1(\rho_c) + 2*P_2(\rho_c) =1$ where
$P_1$ and $P_2$ are the probabilities for propagation and branching 
respectively, introduced before.
We have investigated analytically and numerically 
that the mean-field theory
is insensitive to the exact functional form of the birth-death
probabilities
and avalanches always decay with an exponent $\tau
=1.5$ at the critical density.  A more detailed analysis of
the above equations considering different explicit forms of
$F$ is considered elsewhere \cite{skvlsr}. 

The scaling relation (\ref{taugam}) always holds for systems with open
boundary conditions provided that there is a compact
bulk packing.
This poses an upper limit to the exponent
$\tau$. For non-fractal bulk packings with a smooth free surface,
$\gamma$ cannot be larger
than $1$ and hence $\tau$ cannot be larger than $1.5$.
It is interesting to note that the avalanches decay with the same exponent
as in mean-field theory. However, the reason for
this exponent here is that the avalanches propagate
in a conical region (implied by $\gamma \sim 1$ )
centered around the grain boundary (since, as mentioned before,
avalanches propagate most of the time at the grain boundary).
These facts imply, from the scaling relation 
(\ref{taugam}), that $\tau=1.5$.

Although our results have so far shown
a universal behavior, we identify within the framework of this 
theory at least
one clear instance of the breakdown of this universality. This
has to do with having a very loose packing in the system. If
this is the case, particles can fall large distances in the course of
an avalanche and compactify the system far below. It would then 
not only be
the width of the avalanche at height $h$ which would contribute to the
compactification but some fraction of the {\it whole} avalanche above
$h$.
Such an effect is clearly
not taken into account in Eq.(\ref{couple1}) which hence 
implicitly assumes
that particles only fall short distances.
We thus have to rewrite the above mean-field theory for
a loose system for which the particles can fall large distances. 
We can quantify the above statements by rewriting
Eq.(\ref{couple1}) in the following manner:

\begin{equation}
\rho_h(t+1) -\rho_h(t) =  -\frac{1}{L^2} +
\frac{1}{L^{2}} \int_{s^*}^{\infty} (s-s^{*})^{\alpha} s^{-\tau}ds
\label{couple3}
\end{equation}
where $s^{*}$ is the typical size of an avalanche reaching a height
$h$ and $\alpha$ is a measure of how much of this avalanche
contributes to heights less than $h$. Making a change of variables and 
taking the $s^{*}$ dependence out of the integral, we find that
the relevant scaling relation is now expressed in terms of $s^{*}$ as
$\tau=1+\alpha$.
Since $\alpha \le 1$ (if the total avalanche above $h$ 
contributes then $\alpha=1$)
we find that for systems with very loose
packings the upper bound for the avalanche distribution is now 
$\tau \le 2$ and not $1.5$ as before.

We have checked this by changing the stability condition for
particles to get a much looser packing. In all the cases considered
above, the particles need to be stable in two directions in order not
to fall. We modified  this by 
looking at a system of {\em sticky} particles, in which one downward
contact in either direction suffices for stability.
Repeating the same recycling procedure used throughout the paper we
find in this case a stationary state with a non-compact bulk packing
(Fig.\ref{structure}-right).
For this system, we find
an avalanche distribution in the steady state 
characterized by an exponent $\tau = 1.9 \pm 0.05$ 
(see Fig.\ref{figaval}) out of the range of
validity of Eq.(\ref{taugam}) and in the range of validity of 
the scaling relation predicted by Eq. (\ref{couple3}).

There are several features that it is of interest to investigate 
further.
An instability mechanism for producing structured steady states
has yet to be developed \cite{skvlsr}. Further it would be interesting
to see how these structures coexist with power-law avalanches and
whether finite driving destroys this effect.
In this context it can be seen from Fig. \ref{figaval} that
the big avalanches are enhanced well over the power-law. It could be
of interest to investigate whether this is just a finite size
effect or whether the structures play a role in this \cite{skvlsr}. 
Within the context of this model
we have also studied a system of spherical
particles  ({\it i.e.} crosses with
roughly equal extensions in either direction). This is the case
closest to the one studied in
\cite{sny-ball}  in which the particles are all the same shape. 
We find that though a density plot is not sufficient to spot 
structures, an activity plot (marking how many avalanches pass through
every site over a period of time) shows very distinctly
that there are always long-lived loose regions
where avalanches preferentially propagate with $\tau \sim 1.5$.
It is hence tempting to conclude that this dynamics always results
in long-lived inhomogeneities 
(with easy channels for particle flow) which affect the avalanche
distribution.
Finally, it is interesting to speculate what our results imply for
possible experiments on the phenomenon of internal avalanches. One 
implication might be that a real system subjected to the
continual process of removal and addition of grains will ``fracture''
(as in our model) developing easy regions for particle flow. It would
be very interesting to see whether this is observable.

{\bf Acknowledgements}: HJH, SK, SSM and SR would like to thank CEFIPRA.
In particular, SK and SSM would like to acknowledge financial 
support from CEFIPRA under project no 1508-3/192. 
VL acknowledges financial support under project ERBFMBICT961220.
This work has also been partially supported from the European 
Network-Fractals under contract No. FMRXCT980183.




%

\end{document}